# A Density Functional Study of Atomic Hydrogen Adsorption on Plutonium Layers

M. N. Huda and A. K. Ray*
P.O. Box 19059, Department of Physics, The University of Texas at Arlington
Arlington, Texas 76019

Hydrogen adsorption on δ-Pu (100) and (111) surfaces using the generalized gradient approximation of the density functional theory with Perdew and Wang functionals have been studied at both the spin-polarized level and the non-spin-polarized level. For the (100) surface at the non-spin-polarized level, we find that the center position of the (100) surface is the most favorable site with a chemisorption energy of 2.762eV and an optimum distance of the hydrogen adatom to the Pu surface of 1.07 A. For the spin-polarized (100) surface, the center site is again the preferred site with a chemisorption energy of 3.467eV and an optimum hydrogen distance of 1.13A. For the non-spin-polarized (111) surface, the center position is also the preferred site, but with slightly lower chemisorption energy, namely 2.756eV and a higher hydrogen distance, 1.40 A, compared to the (100) center site. The center site is also the preferred site for the spin-polarized (111) surface, with a chemisorption energy of 3.450eV and a hydrogen distance of 1.42 A. Also, for the spin-polarized calculations, the over all net magnetic moments of the (111) surface changed significantly due to the hydrogen adsorption. The 5f orbitals are delocalized, especially as one approaches the Fermi level. However, the degree of localization decreases for spin-polarized calculations. The coordination numbers have a significant role in the chemical bonding process. Mulliken charge distribution analysis indicates that the interaction of Pu with H mainly takes place in the first layer and that the other two layers are only slightly affected. Work functions, in general, tend to increase due to the presence of a hydrogen adatom.

---

*email: akr@exchange.uta.edu

## A. Introduction

Considerable theoretical efforts have been devoted in recent years to studying the electronic and geometric structures and related properties of surfaces to high accuracy. One of the many motivations for this burgeoning effort has been a desire to understand the detailed mechanisms that lead to surface corrosion in the presence of environmental gases; a problem that is not only scientifically and technologically challenging but also environmentally important. Such efforts are particularly important for systems like the actinides for which experimental work is relatively difficult to perform due to material problems and toxicity. As is known, the actinides are characterized by a gradual filling of the 5f-electron shell with the degree of localization increasing with the atomic number Z along the last series of the periodic table. The open shell of the 5f electrons determines the magnetic and solid-state properties of the actinide elements and their compounds and understanding the quantum mechanics of the 5f electrons is the defining issue in the physics and chemistry of the actinide elements. These elements are also characterized by the increasing prominence of relativistic effects. Studying them can, in fact, help us to understand the role of relativity throughout the periodic table. Narrower 5*f* bands near the Fermi level, compared to 4*d* and 5*d* bands in transition elements, are believed to be responsible for the exotic structure of actinides at ambient condition [1]. The 5f orbitals have properties intermediate between those of localized 4f and delocalized 3d orbitals and, as such, the actinides constitute the "missing link" between the d transition elements and the lanthanides [2]. Thus, a proper and accurate understanding of the actinides will help us understand the behavior of the lanthanides and transition metals as well.

Among the actinides, plutonium (Pu) is particularly interesting in two respects [3-6]. First, Pu has, at least, six stable allotropes between room temperature and melting at atmospheric pressure, indicating that the valence electrons can hybridize into a number of complex bonding arrangements. Second, plutonium represents the boundary between the light actinides, Th to Pu, characterized by itinerant 5f electron behavior, and the heavy actinides, Am and beyond, characterized by localized 5f electron behavior. In fact, the high temperature fcc δ-phase of plutonium exhibits properties that are intermediate between the properties expected for the light and heavy actinides. These unusual aspects of the bonding in bulk Pu are apt to be enhanced at a surface or in an ultra thin film of Pu adsorbed on a substrate, due to the reduced atomic coordination of a surface atom and the narrow bandwidth of surface states. For this reason, Pu surfaces and films and adsorptions on these may provide a valuable source of information about the bonding in Pu.

This work has concentrated on square and hexagonal Pu layers corresponding to the (100) and (111) surfaces of δ-Pu and adsorptions of hydrogen adatoms on such surfaces, using the formalism of modern density functional theory. Although the monoclinic α-phase of Pu is more stable under ambient conditions, there are advantages to studying δ-like layers. First, a very small amount of impurities can stabilize δ-Pu at room temperature. For example, $Pu_{1-x}Ga_x$ has the fcc structure and physical properties of δ - Pu for $0.020 \leq x \leq 0.085$ [7]. Second, grazing-incidence photoemission studies combined with the calculations of Eriksson *et al.* [8] suggest the existence of a small-moment δ-like surface on α-Pu. Our work on Pu monolayers has also indicated the possibility of such a surface [9]. Recently, high-purity ultrathin layers of Pu deposited on Mg were studied by X-ray photoelectron (XPS) and high-resolution valence band (UPS) spectroscopy by Gouder *et al* [10]. They found that the degree of delocalization of the 5f states depends in a very dramatic way on the layer thickness and the itinerant character of the 5f states is gradually lost with reduced thickness, suggesting that the thinner films are δ-like. Localised 5f states, which appear as a broad peak 1.6 eV below the Fermi level, were observed for one monolayer. At intermediate thickness, three narrow peaks appear close to the Fermi level and a comparative study of bulk α-Pu indicated a surface reorganization yielding more localized f-electrons at thermodynamic equilibrium. Finally, it may be possible to study 5f localization in Pu through adsorptions on carefully selected substrates for which the adsorbed layers are more likely to be δ-like than α-like.

The anomalous properties of δ-plutonium have triggered extensive studies on its electronic structures and ground state properties over the years. Different levels and types of theories have been proposed and used to deal with this strongly correlated system. Standard density functional theory (DFT), which works well for the lighter actinides, was found to be inadequate to for the description of some of the ground state properties of δ-Pu [11]. For example, DFT in the local density approximation (LDA) for the electron exchange and correlation effects underestimates the equilibrium volume up to 30% and predicts an approximately four times too large bulk modulus [12-13]. The electronic structure is, in fact, incompatible with photoemission spectra. On the other hand, theories beyond LDA, such as, the self-interaction-corrected (SIC) LDA studied by Petit *et al.* [14] predicted a 30% too large equilibrium volume. Penicaud [15] performed total energy calculations in the local density approximation using fully relativistic muffin-tin orbital band structure method. For δ-Pu, the $5f_{5/2}$ electrons are uncoupled from the s, p and d electrons to reproduce the experimental value of the equilibrium atomic volume. Also an adjustable parameter was introduced to get a better theoretical representaion of δ-Pu. Using 'mixed-level' model, where the energies were calculated at both localized and delocalized 5f configurations, Eriksson *et al.* [16] reproduced reseasonable equilibrium volumes of U, Pu and Am. There have been also attempts to use the LDA+U method, where U is the adjustable Hubbard parameter, to describe the electron correlation within the dynamical mean field theory (DMFT) [17]. The experimental equilibrium δ-Pu volume was reproduced, with U equal to 4 eV.

As is known, the existence of magnetic moments in bulk δ-Pu is a subject of great controversy and significant discrepancies exist between various experimental and theoretical results. To this end, we comment on a few representative works in the literature, partly to mention explicitly some of the controversies. Susceptibility and resistivity data for δ-Pu were published by Meot-Reymond and Fournier [18], which indicated the existence of small magnetic moments screened at low temperatures. This screening was attributed to the Kondo effect. Recent experiments by Curro and Morales [7] of 1.7 percent Ga-doped δ-Pu conducted at temeperatures lower than the proposed Kondo Temperature of 200-300 K showed little evidence for local magnetic moments at the Pu sites. Though there is no direct evidence for magnetic moment, spin-polarized DFT, specifically the generalized-gradient-approximation (GGA) to DFT, has been used by theoreticians, in particular, to predict the magnetic ordering and the ground state properties of δ-Pu. This is partly due to the fact that spin-polarized DFT calculations do predict better agreement with photoemission data. Niklasson *et al*. [19] have presented a first- principles disordered local moment (DLM) picture within the local-spin-density and coherent potential approximations (LSDA+CPA) to model some of the main characteristics of the energetics of the actinides, including δ-Pu. The authors also descibed the failures of the local density approximation (LDA) to describe 5f localization in the heavy actinides, including elemental Pu. The DLM density of states was found to compare well with photoemission on δ-Pu, in contrast to that obtained from LDA or the magnetically ordered AFM configuration. On the other hand, Wang and Sun [20], using the full-potential linearized augmented-plane-wave (FP-LAPW) method within the spin-polarized generalized gradient approximation (SP-GGA) to density functional theory, without spin-orbit coupling, found that that the antiferromagnetic-state lattice constant and bulk modulus agreed better with experimental values than the nonmagnetic values of δ-Pu. Using the fully relativistic linear combinations of Gaussian-type orbitals-fitting function (LCGTO- FF) method within GGA, Boettger [21] found that, at zero pressure, the AFM (001) state was bound relative to the non-magnetic state by about 40 mRy per atom. The lattice constant for the AFM (001) state also agreed better with the experimental lattice constant as compared to the nonmagnetic lattice constant. However, the predicted bulk modulus was significantly larger than the experimental value. Söderlind *et al.* [22], employing the all electron, full-potential-linear-muffin-tin-orbitals (FLMTO) method, predicted a mechanical instability of antiferromagnetic δ-Pu, and proposed that δ-Pu is a 'disordered magnet'. In a more recent study on 5f localization, Söderlind *et al.* showed that 5f-band fractional occupation at 3.7 (68% atoms with itinerant 5f electrons) can predict well the atomic volume and bulk modulus without referring to the magnetic ordering. Wills *et al.* [23] have claimed that there is, in fact, no evidence of magnetic moments in the bulk δ- phase, either ordered or disordered. We also wish to mention that *no detailed information exists in the literature about the magnetic state of the surface of* $\delta$ *-Pu* and our present study including spin polarization on adsorptions on Pu surfaces is a *first step* towards an understanding of atomic chemisorption on Pu surfaces *and* the influence of magnetism on such surfaces. We also note that, as the films get thicker, the complexity of magnetic ordering, if existent, increases and such calculations are quite challenging computationally. Nevertheless, to study the effects of spin polarization on the chemisorption process, our studies have been performed at both the spin-polarized and at the non-spin-polarized levels.

The literature on hydrogen atomic adsorption on Pu surfaces is relatively scarce. Some work has concentrated on hydrogenated or dehydrogenated plutonium due to its importance on plutonium recovery [24]. The interactions of Pu with $H_2$ and $O_2$ have been investigated in several works [25]. Hydrogen also acts to catalyze the oxidation of plutonium [26]. Using the film-linearized-muffin-tin-orbitals (FLMTO) method, Eriksson *et al*. [8] have studied the electronic structure of hydrogen and oxygen chemisorbed on Pu. The slab geometry had the $CaF_2$ structure and the chemisorbed atoms were assumed to have four-fold bridging positions at the surface. The surface behaviors in $PuH_2$ and $PuO_2$ were rather different compared to the surface behavior in pure metallic Pu. For metallic Pu, 5f electrons are valence electrons and show only a small covalent like bonding contribution associated with small 5f to non-5f band hybridization. For the hydride and the oxide, the Pu 5f electrons are well localized and treated as core electrons. Thus, the Pu valence behavior is dominated by the 6d electrons, giving rise to significant hybridization with the ligand valence electrons and covalency. The energy gained when H atoms chemisorbed on the Pu surface was 4.0eV per atom. There are *no other* theoretical studies in the literature on hydrogen adsorption on the Pu surface. In our previous study [27] of water adsorption on $PuO_2$ (110) surface, we have found dissociative adsorption favored over molecular adsorption, with the hydrogen interaction being rather weak. In this work we present our results on hydrogen atom adsorption on Pu (100) and (111) surfaces using the formalism of modern density functional theory.

**B. Computational Details**

As in our previous work [27], all computations reported here have been performed at both the spin unrestricted and the spin restricted generalized gradient approximation (GGA) level [28] of density functional theory (DFT) [29], using the suite of programs DMol3 [30]. This code does not yet allow fully relativistic computations and, as such, we have used the scalar-relativistic approach. In this approach, the effect of spin-orbit coupling is omitted primarily for computational reasons, but all other relativistic kinematic effects, such as mass-velocity, Darwin, and higher order terms are retained. It has been shown [30] that this approach models actinide bond lengths fairly well. We certainly do not expect that the inclusion of the effects of spin-orbit coupling, though desirable, will alter the primary qualitative and quantitative conclusions of this paper, particularly since we are interested in chemisorption energies defined as the difference in total energies. We also note that Landa *et al.* [22] and Kollar *et al.* [31] have observed that inclusions of spin-orbit coupling are not essential for the quantitative behavior of $\delta$ - Pu. Hay and Martin [32] found that one could adequately describe the electronic and geometric properties of actinide complexes without treating spin-orbit effects explicitly. Similar conclusions have been reached by us in our study of water adsorption [27] and of molecular $PuO_2$ and $PuN_2$ [33] and by Ismail *et al*. [34] in their study of uranyl and plutonyl ions. We also note that scalar-relativistic hybrid density functional theory has been used by Kudin *et al.* [35] to describe the insulating gap of $UO_2$, yielding a correct antiferromagnetic insulator.

In DMol3, the physical wave function is expanded in an accurate numerical basis set, and fast convergent three-dimensional integration is used to calculate the matrix elements occurring in the Ritz variational method. For the H atom, a double numerical basis set with polarization functions (DNP) and real space cut-off of 5.0 Å was used. The sizes of these DNP basis set are comparable to the 6-31G** basis of Hehre *et al.* [36]. However, they are believed to be much more accurate than a Gaussian basis set of the same size [30]. For Pu, the outer sixteen electrons ($6s^2$ $6p^6$ $5f^6$ $7s^2$) are treated as valence electrons and the remaining seventy-eight electrons are treated as core. A hardness conserving semi-local pseudopotential, called density functional semi-core pseudo-potential (DSPP), has been used. These norm-conserving pseudo-potentials are generated by fitting all-electron relativistic DFT results and have a non-local contribution for each channel up to $l = 2$, as well as a non-local contribution to account for higher channels. To simulate periodic boundary conditions, a vacuum layer of 30 Å was added to the unit cell of the layers. The k-point sampling was done by the use of the Monkhorst-Pack scheme [37]. The maximum number of numerical integration mesh points available in DMol3 has been chosen for our computations, and the threshold of density matrix convergence is set to $10^{-6}$. All computations have been performed on a Compaq ES40 alpha multi-processor supercomputer at the University of Texas at Arlington.

## C. Results and Discussions

To study hydrogen adsorption on fcc (100) and (111) Pu surfaces, the surfaces are modeled with three layers of Pu at the experimental lattice constant. One of the reasons for choosing the experimental lattice constant comes from the fact that problems were encountered in direct applications of DFT to bulk Pu, as mentioned above, and another reason comes from our wish to simulate the experimental chemisorption process as much as possible. The choice of three layers of Pu is believed to be quite adequate considering that the adatom is not expected to interact with atoms beyond the first three layers. This was also the case for our study of oxygen atom adsorption on plutonium surfaces [38]. Also, recently in a study of quantum size effects in (111) layers of δ- Pu, we have shown that surface energies converge within the first three layers [39]. Due to severe demands on computational resources, the unit cell per layer was chosen to contain two Pu atoms. Thus our three-layer model of the surface contains six Pu atoms. The hydrogen atom, one per unit cell, was allowed to approach the Pu surface along four different symmetrical positions: i) directly on top of a Pu atom (*top* position); ii) on the middle of two nearest neighbor Pu atoms (*bridge* position); iii) in the center of the smallest unit structures of the surfaces (*center* position) and iv) inside the Pu layers (*interstitial* position) (Figure 1). The chemisorption energies are calculated from:

$$E_c = E\ (\text{Pu-layers}) + E\ (\text{H}) - E\ (\text{Pu-layers} + \text{H}) \qquad (1)$$

For the non-spin-polarized case, both E (Pu-layers) and E (Pu-layers + H) were calculated without spin polarization, while for spin-polarized computations, both of these two energies are spin polarized. E (H) is the ground state energy of the hydrogen atom. The chemisorption energies and the equilibrium distances of the H atom from the top Pu layer are given in table 1.

We first comment on the hydrogen adsorption on δ-Pu (100) surface in the square symmetry. The chemisorption energies as a function of the separation distances of H atom from the top layer are shown in figures 2(a–h). In non-spin polarized calculations, the center site, the most symmetrical site, is found to be the most favorable chemisorption site with a chemisorption energy of 2.762 eV, followed by the bridge, interstitial and top sites, with chemisorption energies of 2.478 eV, 2.079 eV and 1.750 eV, respectively. The top site has the lowest chemisorption energy for H adsorption, whereas for oxygen adsorption the interstitial site was found to have the lowest energy [38]. For the center position, the distance of the hydrogen atom from the top layer has the lowest value of 1.070 Å, with the four Pu atoms at the corners of the square being 2.395 Å apart. For the bridge position, the next most stable chemisorption position, the distance of the hydrogen atom from the surface is 1.500 Å, and the nearest H-Pu distance is 2.132 Å. For the top position, the normal distance is 1.990 Å. In view of the above picture of distance versus chemisorption energy, it is evident that the most favorable site for chemisorption is influenced by the coordination of the hydrogen atom with the Pu atoms. For the center site, the coordination number is four as compared with a coordination number of one for the top site. Inclusion of spin polarization changes the chemisorption energies significantly, the changes being from 0.367eV to 0.705eV, but the optimum distances remain about the same. The center site is again found to be the most favorable site. Mulliken population analysis [40] indicates (table 2a) that for the center position, there is less charge transfer to the hydrogen atom compared to the other sites. We also note that for the top position, the Pu atom directly below the hydrogen atom is negatively charged, while the surrounding Pu atoms are slightly positively charged. The second and third-layer charge distribution for the center, bridge and top position are least affected by the hydrogen adsorption on Pu layers. In all three cases, chemisorbed hydrogen atom is very slightly negatively charged; hence, unlike oxygen adsorption on Pu surfaces [38], ionic bonding does not play a major part here. For non-spin polarized interstitial sites, we found two equivalent sites, not at the center of the fcc cell, but one at 2.125 Å and the other at 2.155 Å below the top layer Pu atom, of equal chemisorption energy, symmetrically placed above and below the center (figure 2(g)). The chemisorption energy for H in these interstitial positions is 2.079 eV, lower than the energies for the center and bridge positions. Only for this site the hydrogen atom is slightly positively charged. The four nearest Pu atoms in the second layer are also positively charged, and thus push the hydrogen atom towards the negatively charged first layer. The spin polarized interstitial sites give almost similar results, with two equivalent positions for hydrogen at 2.115 Å and 2.165 Å below the top layer, with chemisorption energy of 2.446 eV. However, inclusion of spin polarization did affect the chemisorption curves as is evident from comparing figures 2(g) and 2(h). The magnetic effect made the center of the fcc cell relatively unstable for hydrogen adsorption, with an energy-hill of about 0.5 eV.

We next consider the (111) surface of δ-Pu. The smallest unit of this surface is an equilateral triangle and the same sites as mentioned above are the most symmetrical distinguishable sites. Figure 1 shows the different chemisorption sites and figures 3(a - h) shows the variation of chemisorption energies with adatom distances to the surfaces. Here the center position is the center of the triangle, which is also the most favorable site with hydrogen chemisorption energy of 2.756 eV with a distance to the surface of 1.400 Å. This is followed by the bridge site with a chemisorption energy of 2.574 eV, with a distance from the surface of 1.530 Å. For the top site the chemisorption energy is 1.848 eV and the distance is 1.970 Å. We note that, except for the center position for the (111) surface, the chemisorption energies are consistently higher by about 0.1eV compared to the energies for the (100) surface. This is partly due to the difference in coordination numbers. The slightly lower chemisorption energy for the center site of the (111) surface might be due to the fact that in the (111) surface for the center position, the coordination number is three, whereas for the (100) surface this is four. For the top and the bridge site the coordination number is the same for both surfaces. Inclusion of spin polarization changed the chemisorption energies significantly, the changes being from 0.642eV to 0.694eV. But again, like the (100) surface, optimum distances of the Pu atom to the surface remain about the same. The interstitial site was directly 1.180 Å below the center of the equilateral triangle. The chemisorption energy was 2.194 eV, which is higher than the top site chemisorption energy. For the spin-polarized interstitial site the chemisorption energy is 2.871 eV, and the hydrogen atom is 1.21 Å below the top layer.

As described above, unlike oxygen adsorption on both (100) and (111) Pu surfaces, interstitial sites have higher chemisorption energies than the top sites. We also observe that for both (100) and (111) surfaces, except for positions of the adatom below the surface, the following inequality between the adatom distance from the surface (r) and the corresponding chemisorption energy (C.E.) holds true:

r (center) < r (bridge) < r (top)

C.E. (center) > C.E. (bridge) > C.E. (top)

This implies, as is expected, that the highest chemisorption energy is obtained when the adatom is nearest to the surface. Population analysis (table 3) also indicates, e.g., for the (111) surface, that the charges of the first layers are significantly modified by the presence of the hydrogen atom, whereas those of the second and third layers remained almost the same (except for the interstitial site). Thus the chemisorption activities mainly take place on the first layer with a smaller contribution from the second layer, and the effects decay quickly, even as early as in the third layer. Except for the (100) top site and spin polarized bridge site, where some of the Pu atoms are slightly positively charged, for all the other sites the first and third layers are negatively charged, and the second layer is positively charged. This is true for the bare Pu layers. However, for the interstitial sites, as expected, charge distributions of all the three layers are affected.

To further study the effects of spin-polarization, we have listed in table 1 the spin magnetic moments for different chemisorption sites of the hydrogen adatom. As mentioned before, inclusion of spin polarization with GGA improves the description of bulk δ-Pu. In our calculations, the net spin magnetic moments of bare plutonium layers were 1.47 $\mu_B$/atom and 2.00 $\mu_B$/atom for the 3-layer (100) and (111) surfaces, respectively. Adsorption of a hydrogen atom on a plutonium surface increases the magnetic moments of the (100) surface, except for the top position, and reduces the magnetic moments for the (111) surface significantly. In fact, the moments of the hydrogen atom adsorbed on (111) surfaces are practically negligible. The center site for the (100) surface with the highest chemisorption energy was found to have a spin magnetic moment of 1.720 $\mu_B$/atom, as compared to a very small value of 0.002 $\mu_B$/atom for the center site for the (111) surface. For both surfaces, spin moments have alternating behavior. However, for both the (100) and the (111) surfaces, the spin ordering is affected in a different manner by hydrogen adsorption; consequently, the first layer magnetic moments (the layer on which hydrogen atom approaches) differs significantly. For example, for the center site the magnetic moment of the two atoms in the unit cell of the first layer of the (100) surface is 5.66 $\mu_B$/atom, whereas for the (111) surface the moment is 0.02 $\mu_B$/atom. However, given the fact that the spin magnetic moment of atomic plutonium is 6 $\mu_B$ and the Pu atoms on both the surfaces have individual spin magnetic moment of more than 5 $\mu_B$, we may infer that the Pu atoms almost preserve their atomic behavior. The reason for the significant reduction of spin magnetic moment for the (111) surface due to the presence of a hydrogen atom is that, after the chemisorption of hydrogen atom, the spins on the plutonium slab assume the anti-parallel ordering, i.e., two Pu atoms on each surface have opposite spins, and thereby the net spin magnetic moment is reduced. For all the cases, the chemisorbed hydrogen atom has a very low spin magnetic moment. For the spin-polarized surfaces, the hydrogen adatom, which has a magnetic moment of 1 $\mu_B$ before adsorption, interacts with the local magnetic field of the Pu surfaces. This contributes to the changes in chemisorption energies when spin polarization is considered

A study of the energy levels of the Pu layers before hydrogen adsorption indicates that while the 12 6s and 36 6p electrons are localized, some of the 5f electrons are localized and some tend to be not localized, in agreement with some previous studies [15,22] and in disagreement with others [8]. The degree of localization decreases as one approaches the Fermi level, indicating the nature of the 5f electrons. Around the Fermi level, the 5f electrons are largely delocalized. We do find hybridization of the Pu 7s electrons with the 6d electrons, indicating that the Pu valence behavior might be dominated by the 6d electrons, in agreement with Eriksson *et al* [8]. For the bare (100) Pu layers, it was found that the top of the 5f band is 0.235 eV below the Fermi level. For the (111) surface this is 0.286 eV. For the corresponding spin-polarized cases these values are 0.414 eV and 0.384 eV, respectively, implying that the spin-polarized calculation leads to more localized 5f orbitals. The Fermi surface is basically formed by the 6d-7s hybridized orbitals. For the (100) surface, the energy gap between the 6s and 6p bands is 24.506 eV, and that between the 6p and 5f bands is 15.407 eV (22.928 eV and 14.028 eV for the spin polarized case). For the (111) surface these gaps are 24.638 eV and 15.441 eV in the non-spin polarized case and for the spin-polarized case, these values are 22.388 eV and 14.551 eV, respectively. Upon the adsorption of a hydrogen atom on the Pu surface, the hydrogen 1s orbital hybridizes with the lower end of the Pu 5f orbitals. For the center position at the (100) surface, the most favorable site, the difference between the 5f orbitals and the Fermi energy increases by only 0.001 eV. For the (111) center position, the difference is increased to 0.400 eV. However, incorporating spin polarization leads to a different picture. With spin polarization, after the adsorption of hydrogen, the difference between the top of the 5f band and the Fermi energy decreases for both the (100) and the (111) surfaces compared to bare plutonium film. The differences are 0.403 eV and 0.288 eV for the (100) and the (111) surfaces, respectively.

In table 4, we tabulate the change in work functions due to the hydrogen adsorption on the Pu surfaces. Hydrogen chemisorbed Pu surfaces have higher work functions than pure Pu surfaces. The changes in work function are less for hydrogen adsorption compared to those for oxygen adsorption on the Pu surface [38]. The change in surface dipole moment due to the presence of hydrogen, which is evident from the Mulliken charge distribution, changes the work function of the surface. The increase in work function is minimum for the (100) center position and highest for the (111) top position. Work function changes are higher for the (111) surface than for the (100) surface. Spin polarized and non-spin polarized calculations gave similar results. However hydrogen adsorbed in the interstitial sites lower the work functions, and for the (111) interstitial sites, the change of the work function is only 0.082 eV (0.078 eV for spin-polarized case).

**D. Conclusions**

In summary, we have studied hydrogen adsorption on δ-Pu (100) and (111) surfaces using the generalized gradient approximation of the density functional theory with Perdew and Wang functionals at both the spin-polarized and the non-spin-polarized levels. For the (100) surface both at the non-spin-polarized and spin-polarized levels, we find that the center position of the H adatom is the most favorable site with chemisorption energies of 2.762eV and 3.467eV, respectively. The corresponding energies for the (111) surface are 2.756eV and 3.450eV, respectively. The net spin magnetic moments of the surfaces change significantly upon hydrogen adsorption, specifically for the (111) surface. The 5f orbitals are delocalized, especially as one approaches the Fermi level. However, the degree of localization decreases for spin-polarized calculations. The coordination numbers have a significant role in the chemical bonding process. Mulliken charge distribution analysis indicates that the interaction of H with Pu mainly takes place with the first layer and that the other two layers are only slightly affected. Work functions, in general, tend to increase due to the presence of a hydrogen adatom.

**Acknowledgements**

The authors acknowledge very useful comments from both referees. This work is supported by the Chemical Sciences, Geosciences and Biosciences Division, Office of Basic Energy Sciences, Office of Science, U. S. Department of Energy (Grant No. DE-FG02-03ER15409) and the Welch Foundation, Houston, Texas (Grant No. Y-1525).

Table 1. Chemisorption energies (in eV) and distances of the H atom (in Å) from the Pu surfaces for different positions.

| Surface | Sites | Non-spin polarization | | Spin-Polarized | | Magnetic Moment (in $\mu_B$/atom) |
|---|---|---|---|---|---|---|
| | | Chemisorption Energy (eV) | Distances (Å) | Chemisorption Energy (eV) | Distances (Å) | |
| (100) | Top | 1.750 | 1.99 | 2.122 | 2.03 | 0.206 |
| | Bridge | 2.478 | 1.50 | 3.141 | 1.56 | 1.720 |
| | Center | 2.762 | 1.07 | 3.467 | 1.13 | 1.720 |
| | Interstitial | 2.079 | 2.13 | 2.446 | 2.12 | 1.960 |
| (111) | Top | 1.848 | 1.97 | 2.517 | 2.01 | 0.029 |
| | Bridge | 2.574 | 1.53 | 3.216 | 1.57 | 0.021 |
| | Center | 2.756 | 1.40 | 3.450 | 1.42 | 0.002 |
| | Interstitial | 2.194 | 1.18 | 2.871 | 1.21 | 0.020 |

Table 2(a). Mulliken charge distributions for different chemisorption sites for the Pu (100) surface. Numbers in the first column are the charge distribution of the Pu three layers without hydrogen atom. The other four columns are the different chemisorption sites. NSP indicate no spin polarization and SP is with spin polarization.

| | Layers | without H | Top | Bridge | Center | Interstitial |
|---|---|---|---|---|---|---|
| NSP | H-atom | × | -0.091 | -0.072 | -0.030 | 0.093 |
| | 1st layer | -0.141 | -0.194 | -0.065 | -0.084 | -0.245 |
| | | -0.141 | 0.004 | -0.065 | -0.098 | -0.189 |
| | 2nd layer | 0.281 | 0.279 | 0.253 | 0.271 | 0.195 |
| | | 0.281 | 0.282 | 0.253 | 0.255 | 0.357 |
| | 3rd layer | -0.141 | -0.147 | -0.152 | -0.157 | -0.115 |
| | | -0.141 | -0.133 | -0.152 | -0.156 | -0.097 |
| SP | H-atom | × | -0.137 | -0.137 | -0.102 | 0.011 |
| | 1st layer | -0.121 | -0.149 | 0.007 | -0.033 | -0.146 |
| | | -0.137 | 0.037 | 0.007 | -0.005 | -0.169 |
| | 2nd layer | 0.240 | 0.261 | 0.203 | 0.222 | 0.235 |
| | | 0.276 | 0.243 | 0.203 | 0.180 | 0.345 |
| | 3rd layer | -0.121 | -0.106 | -0.128 | -0.130 | -0.130 |
| | | -0.137 | -0.149 | -0.128 | -0.133 | -0.154 |

Table 2(b). Mulliken charge distributions for different chemisorption sites for the Pu (111) surface. Numbers in the first column are the charge distribution of the Pu three layers without hydrogen atom. The other four columns are the different chemisorption sites. NSP indicate no spin polarization and SP is with spin polarization.

| | Layers | without O | Top | Bridge | Center | Interstitial |
|---|---|---|---|---|---|---|
| NSP | H-atom | × | -0.051 | -0.028 | -0.050 | 0.122 |
| | 1st layer | -0.169 | -0.059 | -0.195 | -0.087 | -0.238 |
| | | -0.169 | -0.213 | -0.015 | -0.095 | -0.198 |
| | 2nd layer | 0.338 | 0.336 | 0.304 | 0.321 | 0.342 |
| | | 0.338 | 0.326 | 0.288 | 0.274 | 0.346 |
| | 3rd layer | -0.169 | -0.166 | -0.179 | -0.183 | -0.194 |
| | | -0.169 | -0.170 | -0.175 | -0.181 | -0.179 |
| SP | H-atom | × | -0.102 | -0.087 | -0.106 | 0.053 |
| | 1st layer | -0.160 | -0.052 | -0.127 | -0.045 | -0.190 |
| | | -0.160 | -0.165 | -0.013 | -0.078 | -0.198 |
| | 2nd layer | 0.320 | 0.336 | 0.300 | 0.321 | 0.369 |
| | | 0.320 | 0.313 | 0.281 | 0.276 | 0.343 |
| | 3rd layer | -0.160 | -0.158 | -0.185 | -0.187 | -0.193 |
| | | -0.160 | -0.172 | -0.168 | -0.180 | -0.185 |

Table 3. Work function changes due to the hydrogen chemisorption on Pu surfaces.

| | Change in work function in eV | | | |
|---|---|---|---|---|
| Sites | (100) surfaces | | (111) surfaces | |
| | NSP | SP | NSP | SP |
| Top | 0.753 | 0.876 | 0.805 | 0.888 |
| Bridge | 0.330 | 0.411 | 0.406 | 0.482 |
| Center | 0.142 | 0.174 | 0.313 | 0.337 |
| Interstitial | –0.118 | –0.103 | –0.082 | –0.078 |

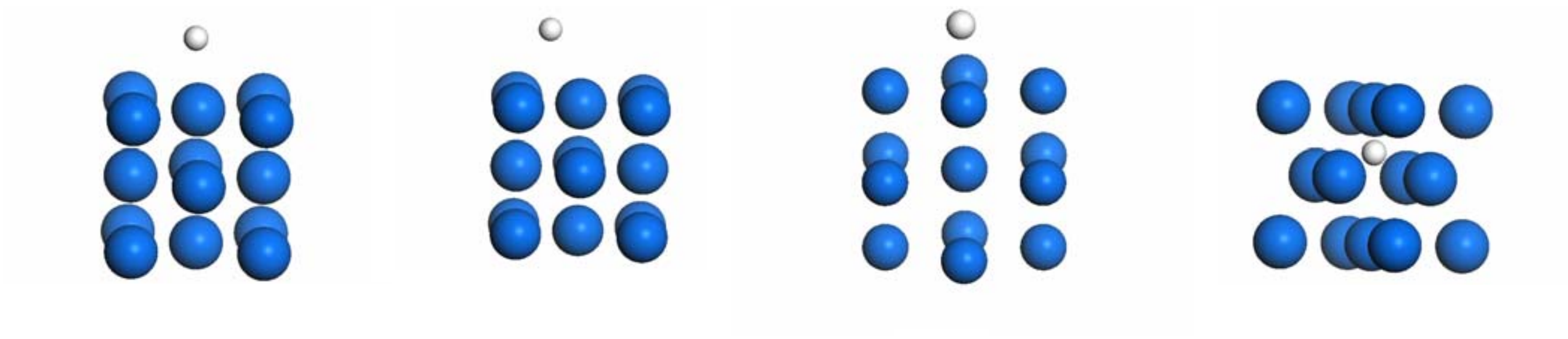

(a) (b) (c) (d)

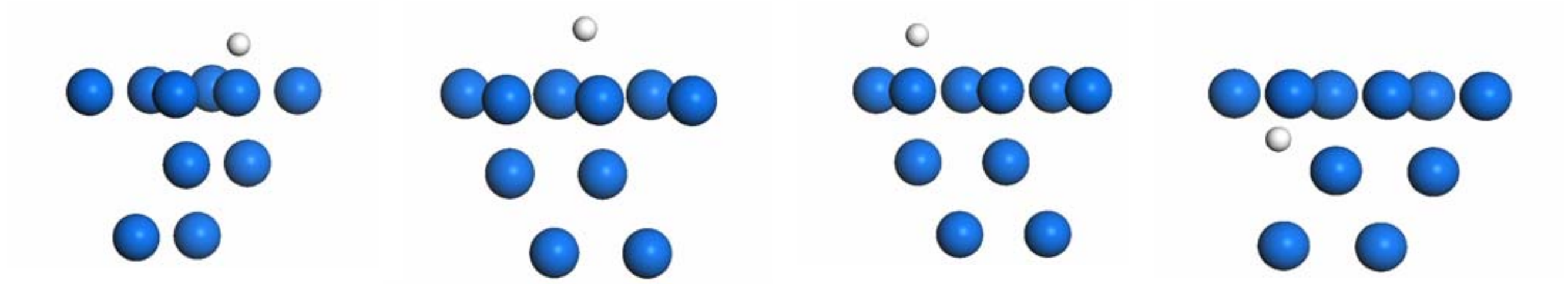

(a) (b) (c) (d)

Figure 1. Different chemisorption sites for (100) surface in the first row and (111) surface in the second row: (a) *top*, (b) *bridge,* (c) *center* and (d) *interstitial* sites.

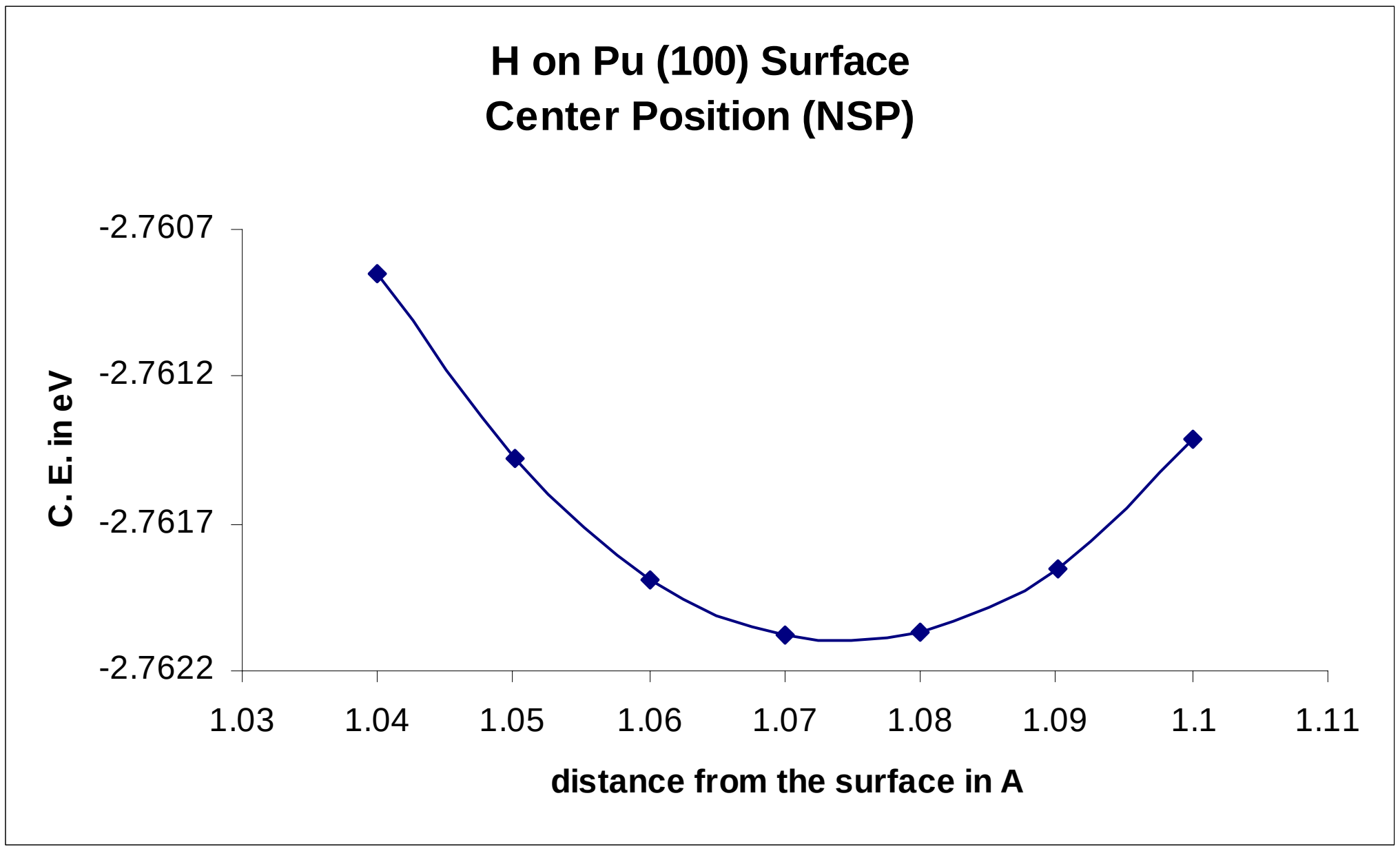

Figure 2(a). Chemisorption energy versus the hydrogen adatom distance from the Pu (100) surface at the center position without spin polarization.

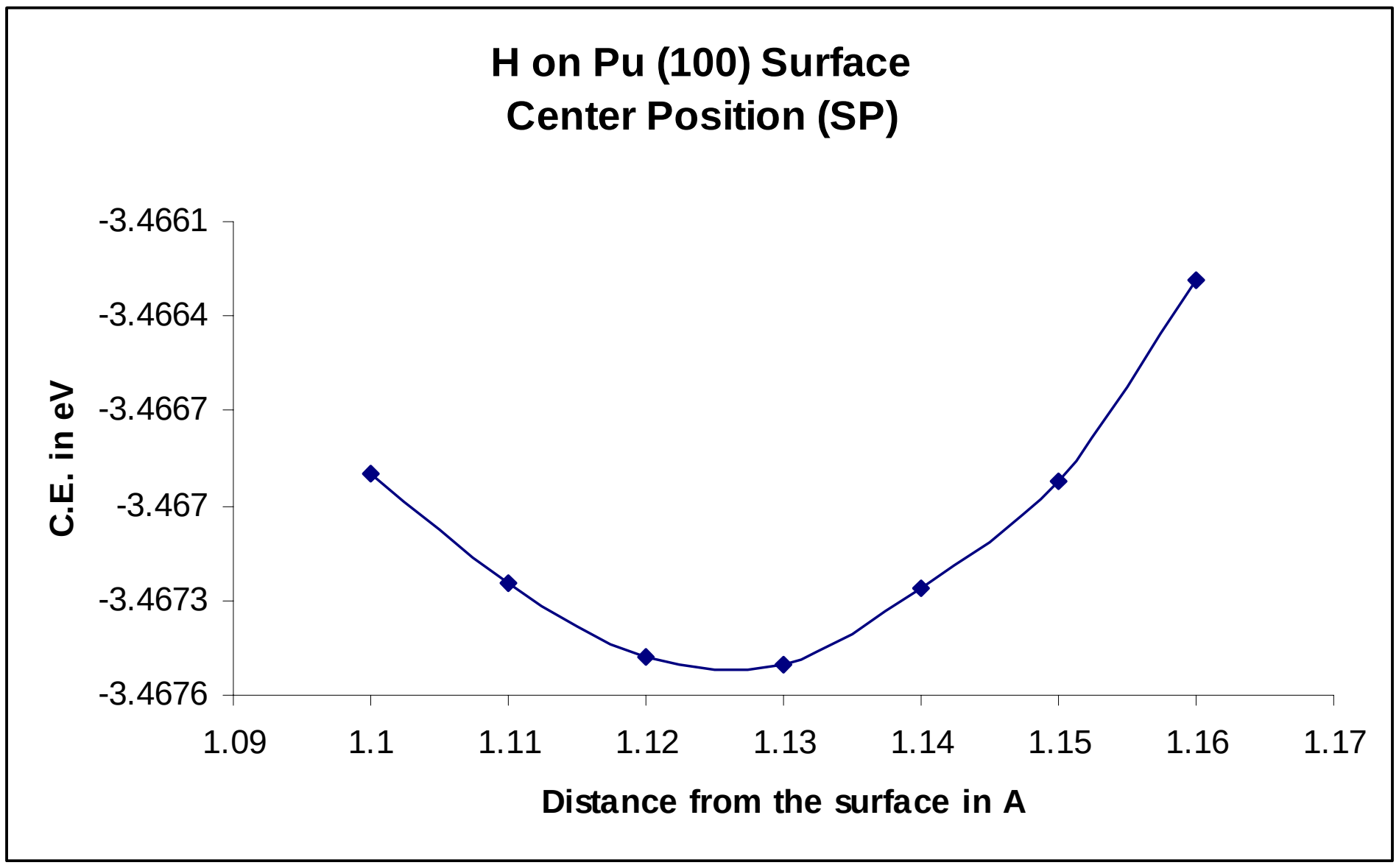

Figure 2(b). Chemisorption energy versus the hydrogen adatom distance from the Pu (100) surface at the center position with spin polarization.

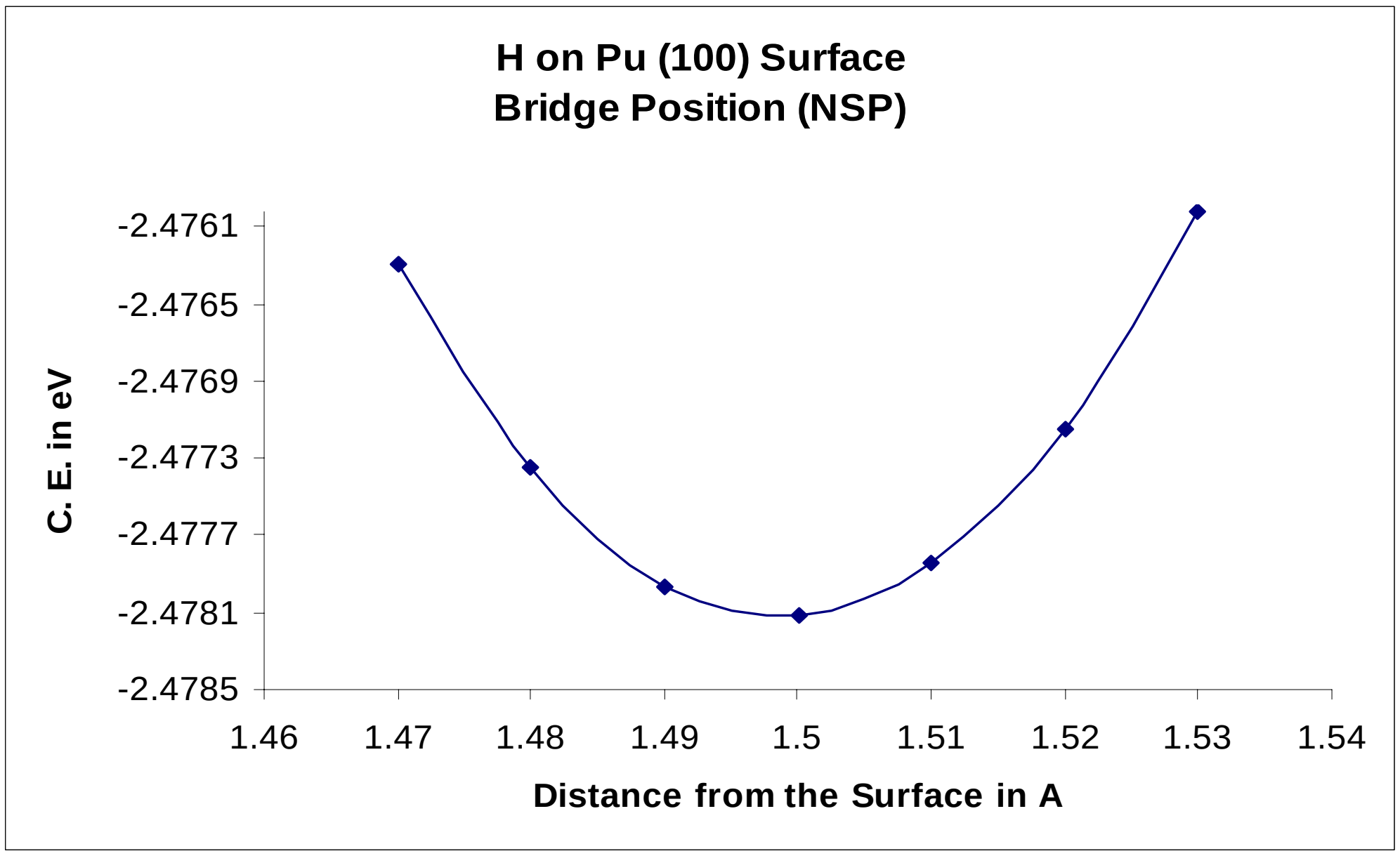

Figure 2(c). Chemisorption energy versus the hydrogen adatom distance from the Pu (100) surface in the bridge position without spin polarization.

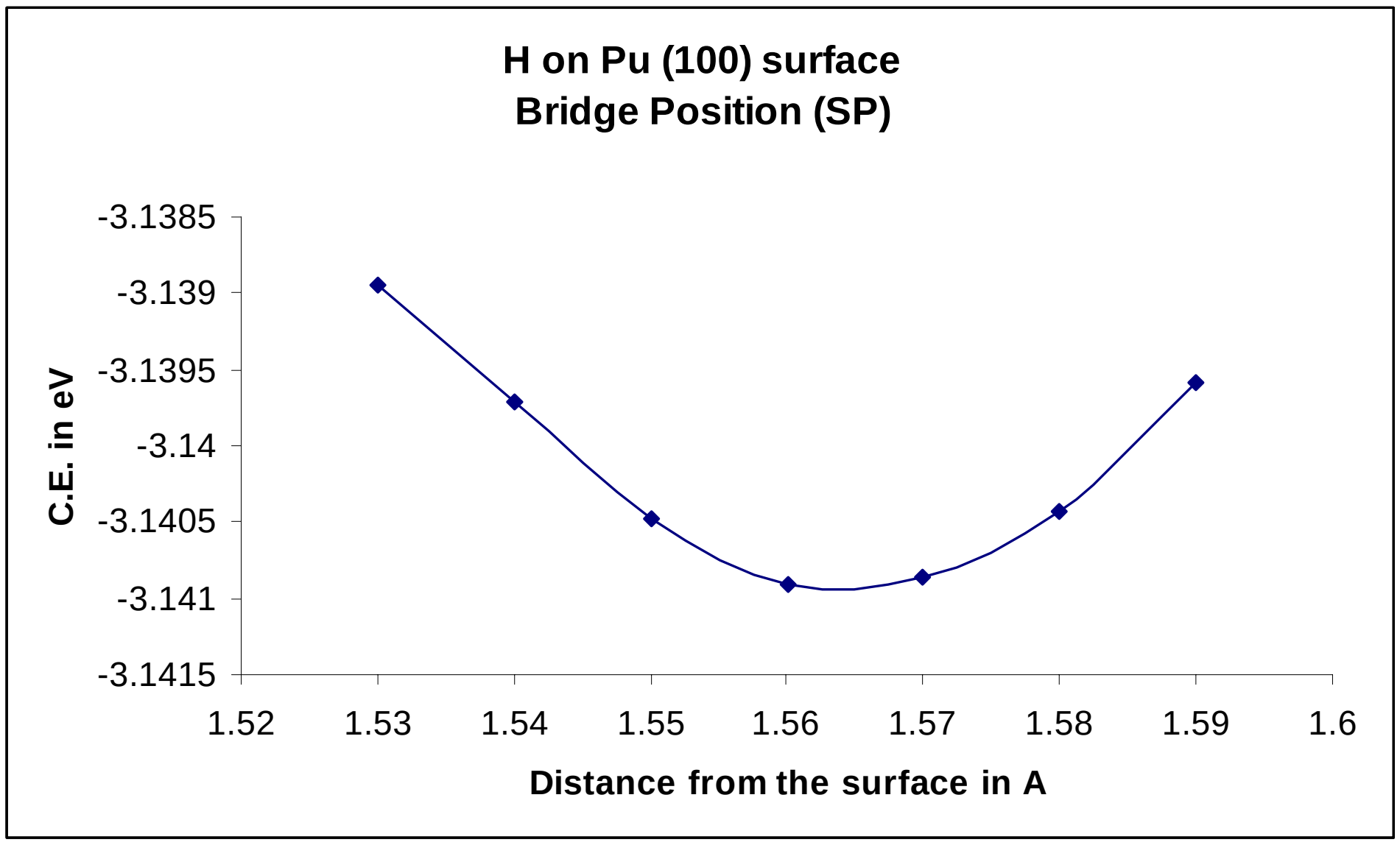

Figure 2(d). Chemisorption energy versus the hydrogen adatom distance from the Pu (100) surface in the bridge position with spin polarization.

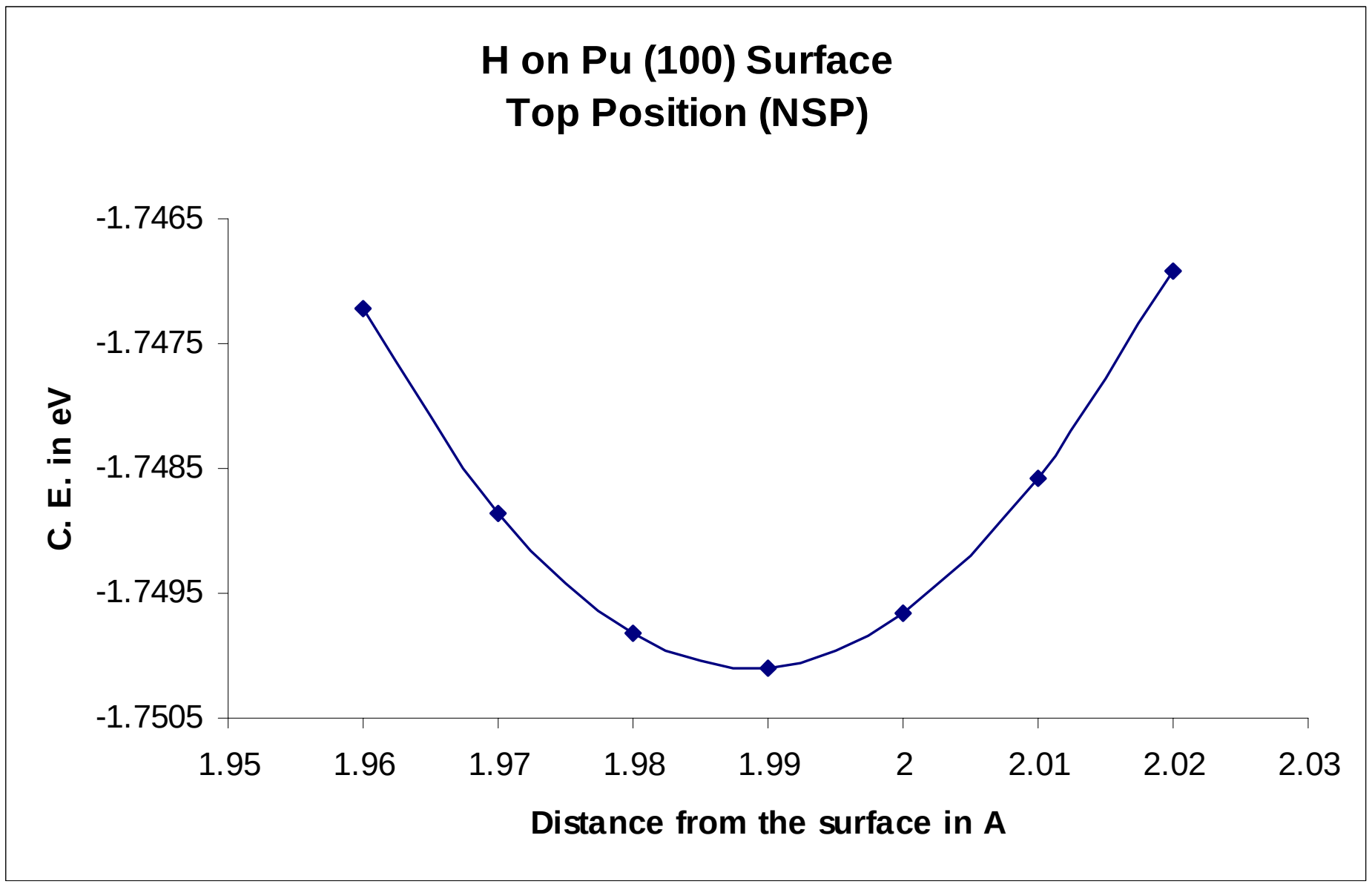

Figure 2(e). Chemisorption energy versus the hydrogen adatom distance from the Pu (100) surface in the top position without spin polarization.

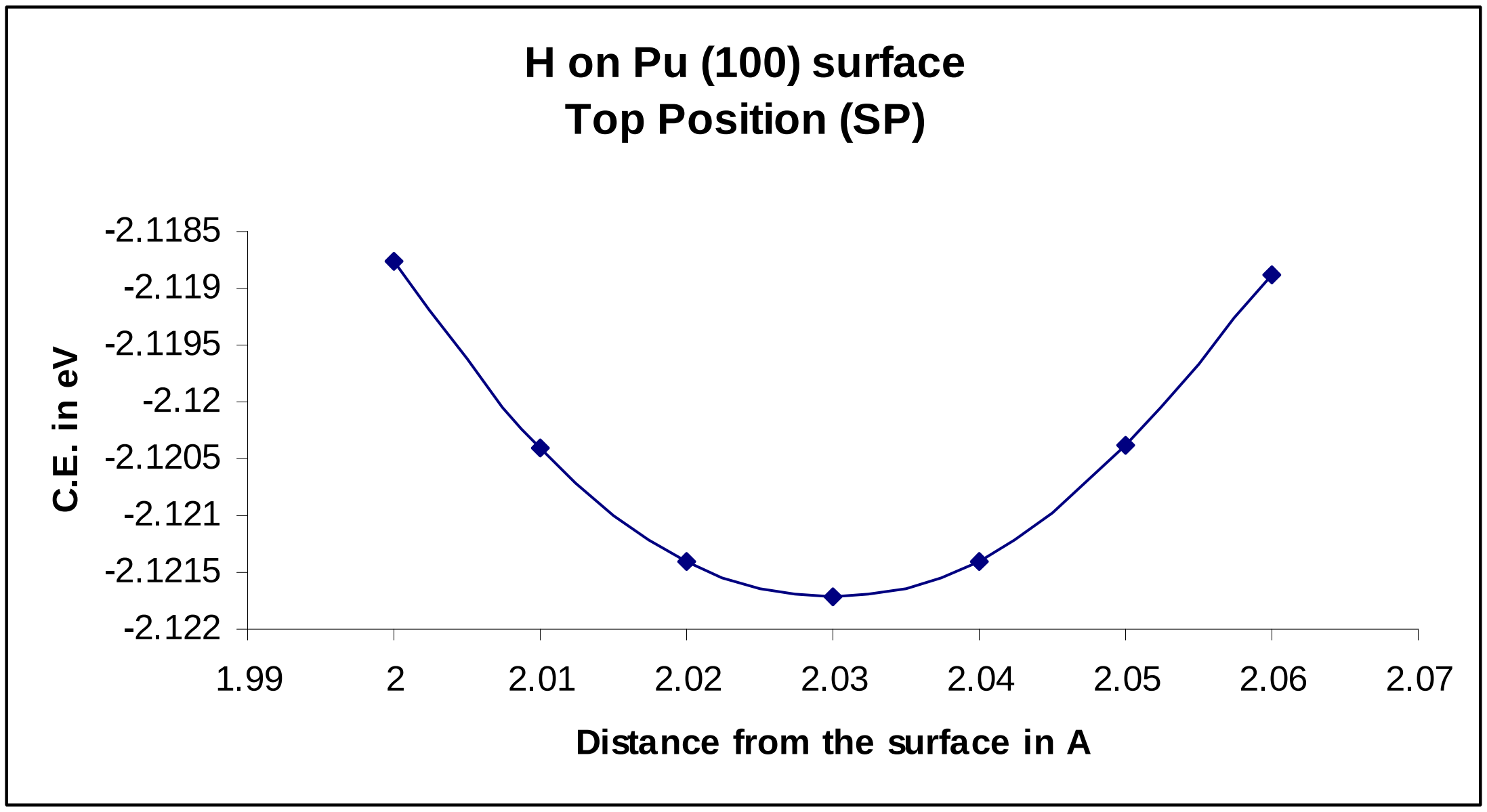

Figure 2(f). Chemisorption energy versus the hydrogen adatom distance from the Pu (100) surface in the top position with spin polarization.

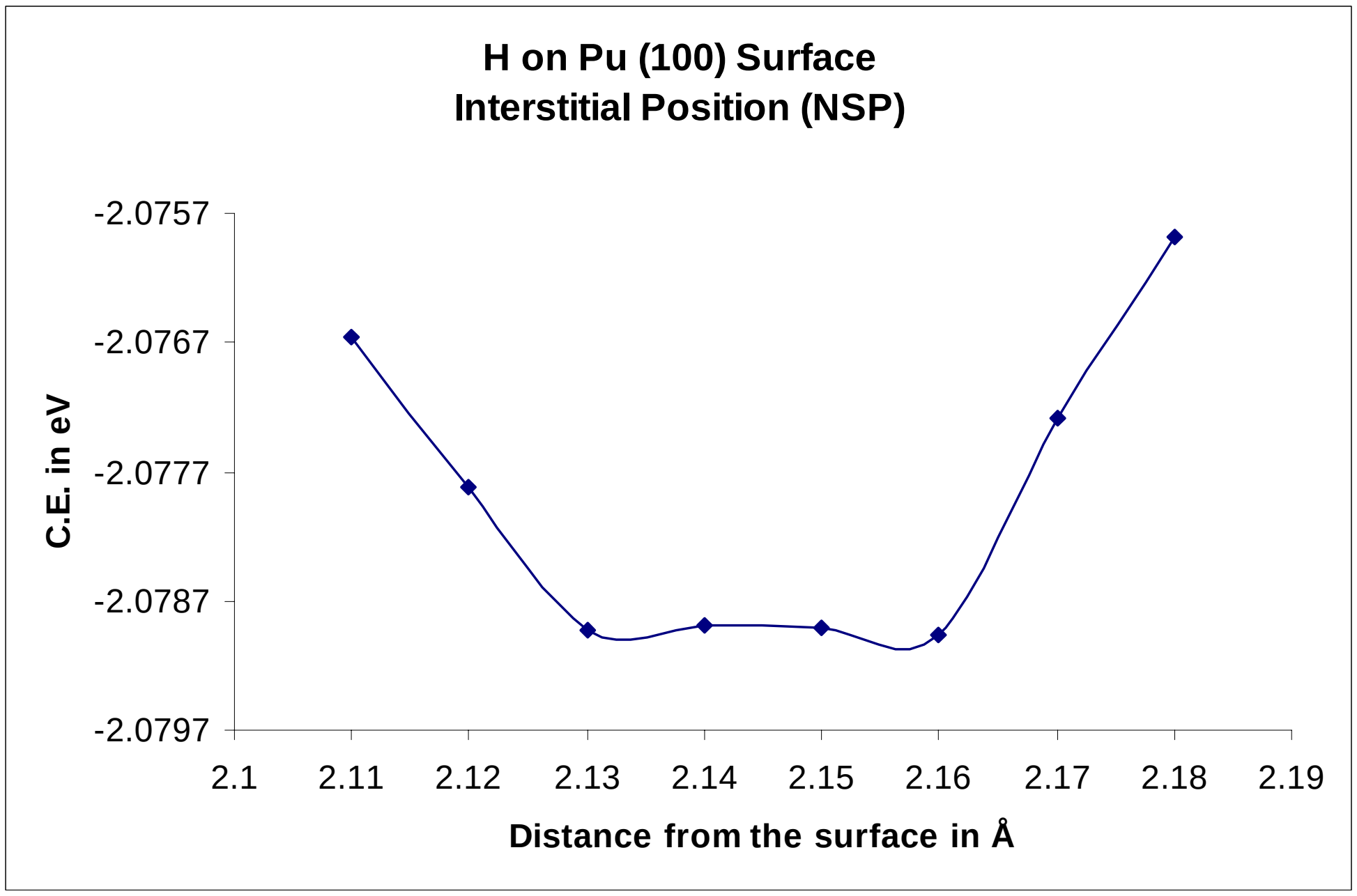

Figure 2(g). Chemisorption energy versus the hydrogen adatom distance from the Pu (100) surface in the interstitial position without spin polarization.

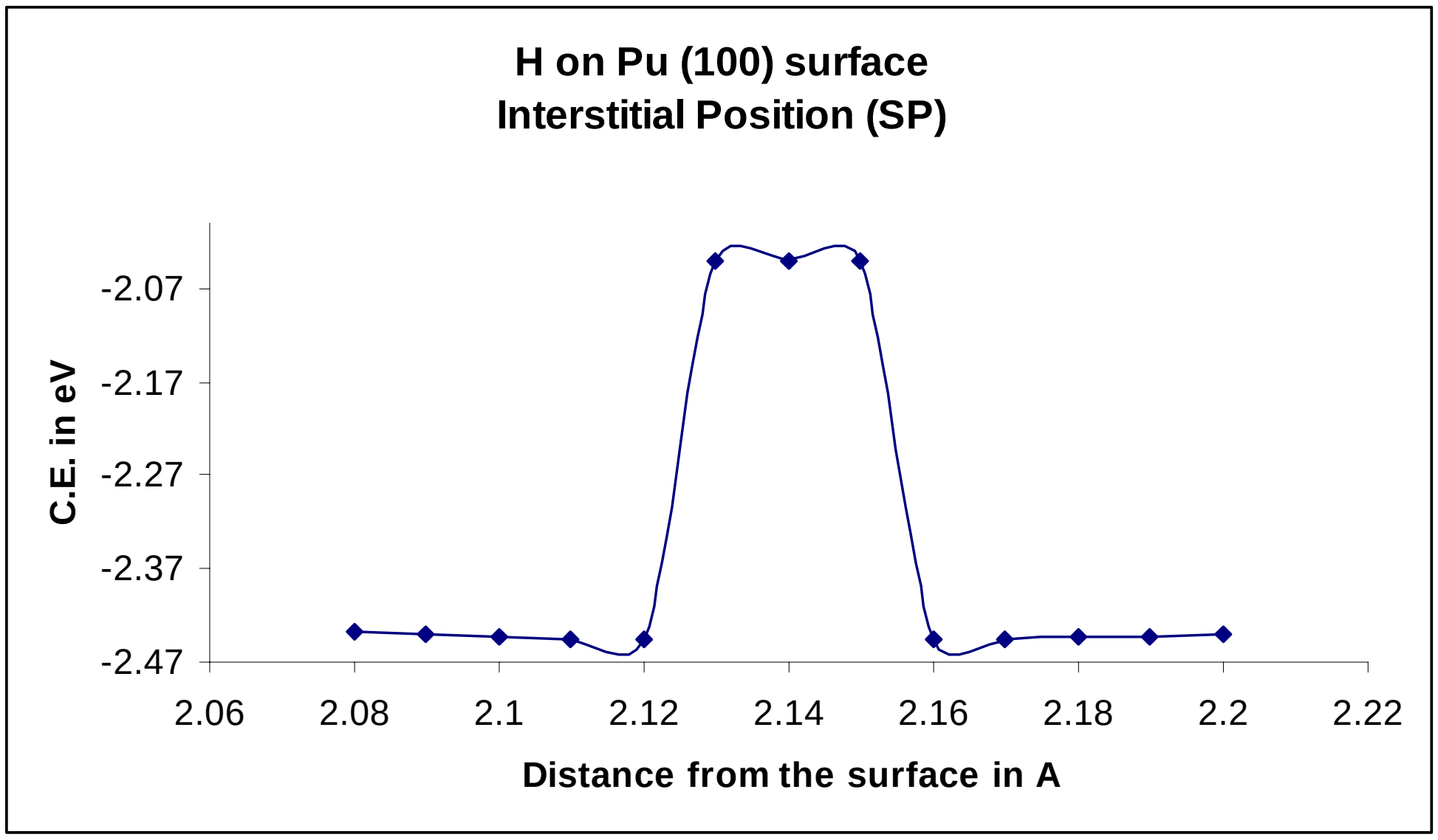

Figure 2(h). Chemisorption energy versus the hydrogen adatom distance from the Pu (100) surface in the interstitial position with spin polarization.

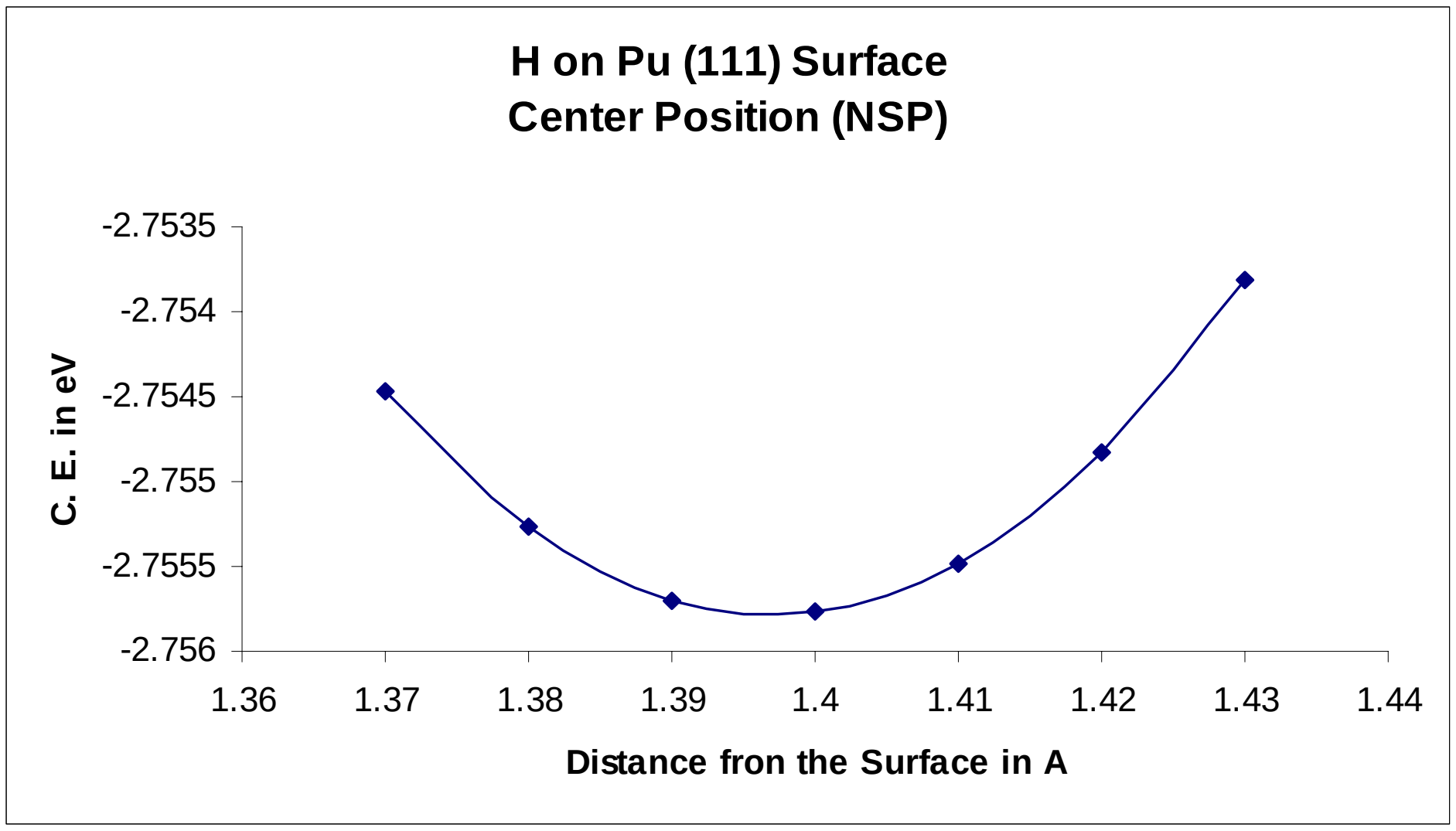

Figure 3(a). Chemisorption energy versus the hydrogen adatom distance from the Pu (111) surface in the center position without spin polarization.

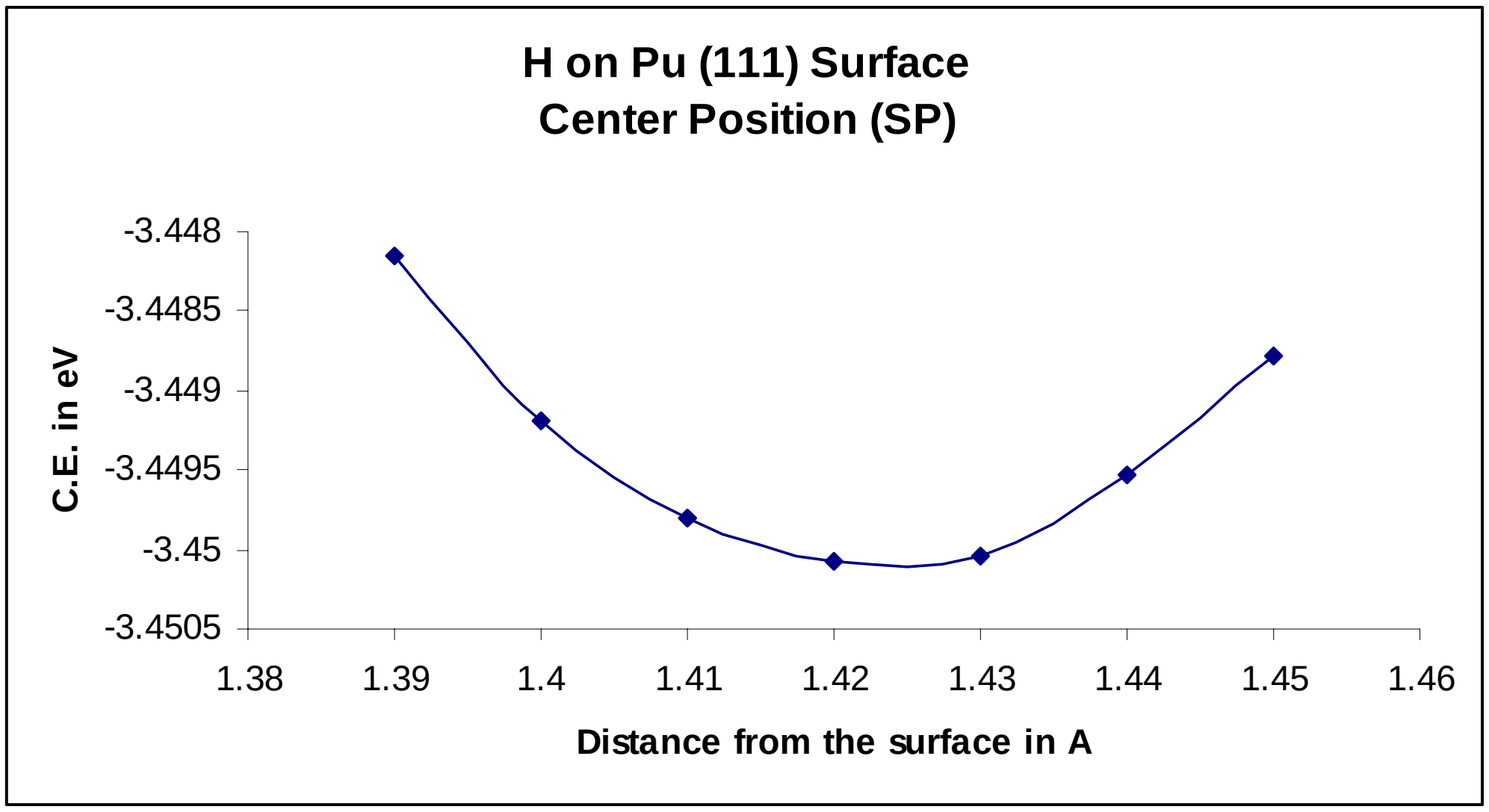

Figure 3(b). Chemisorption energy versus the hydrogen adatom distance from the Pu (111) surface in the center position with spin polarization.

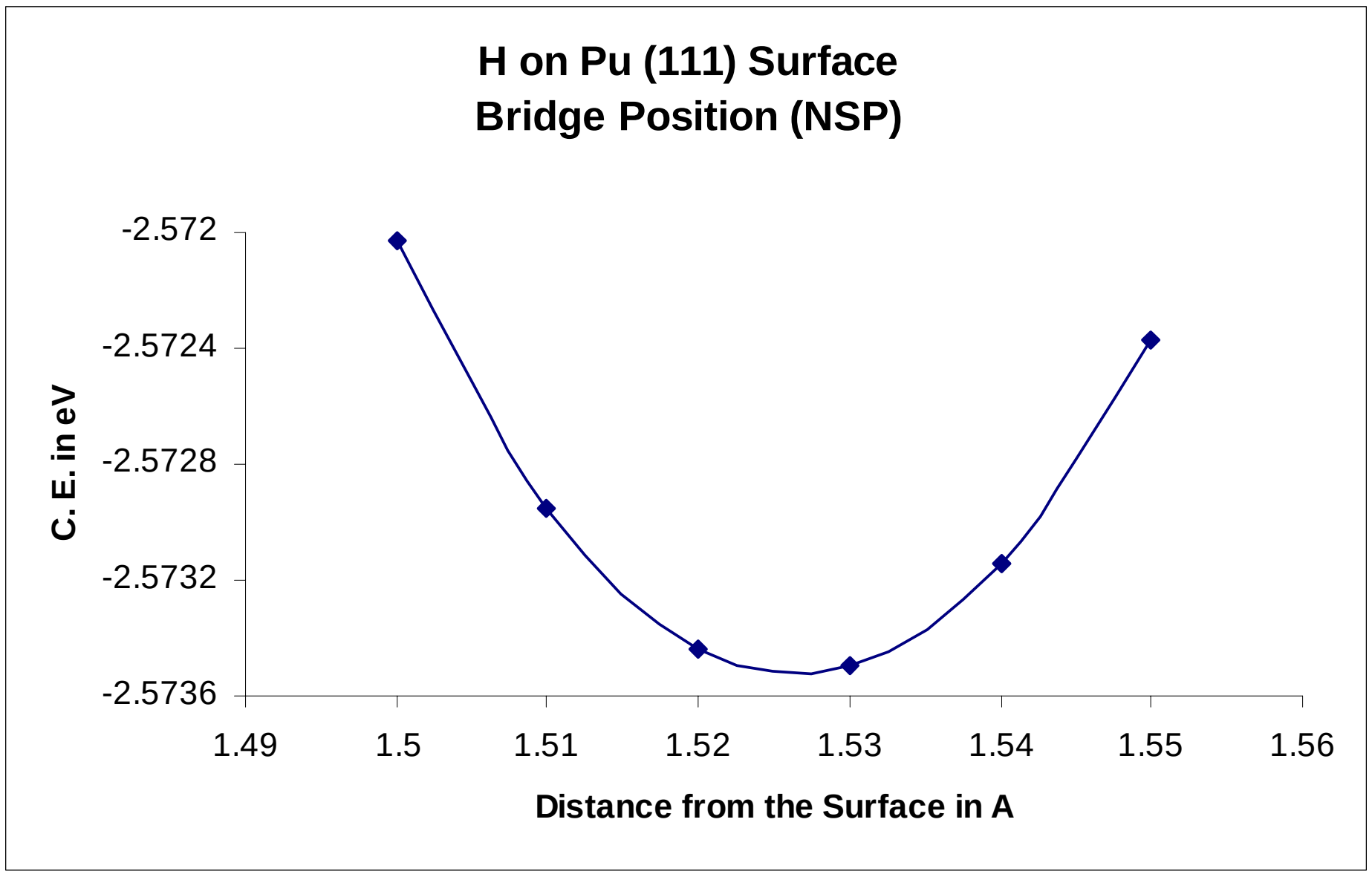

Figure 3(c). Chemisorption energy versus the hydrogen adatom distance from the Pu (111) surface in the bridge position without spin polarization.

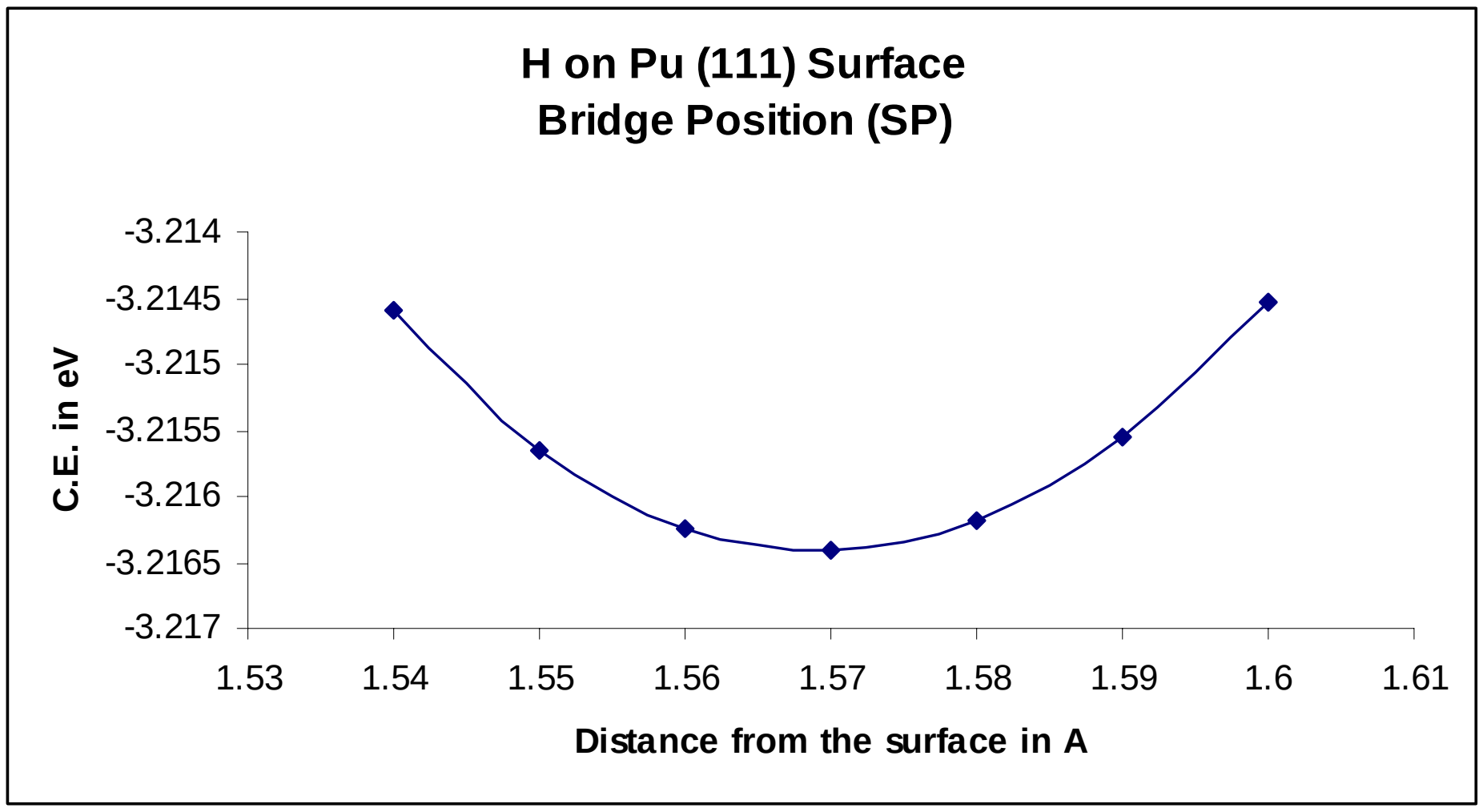

Figure 3(d). Chemisorption energy versus the hydrogen adatom distance from the Pu (111) surface in the bridge position with spin polarization.

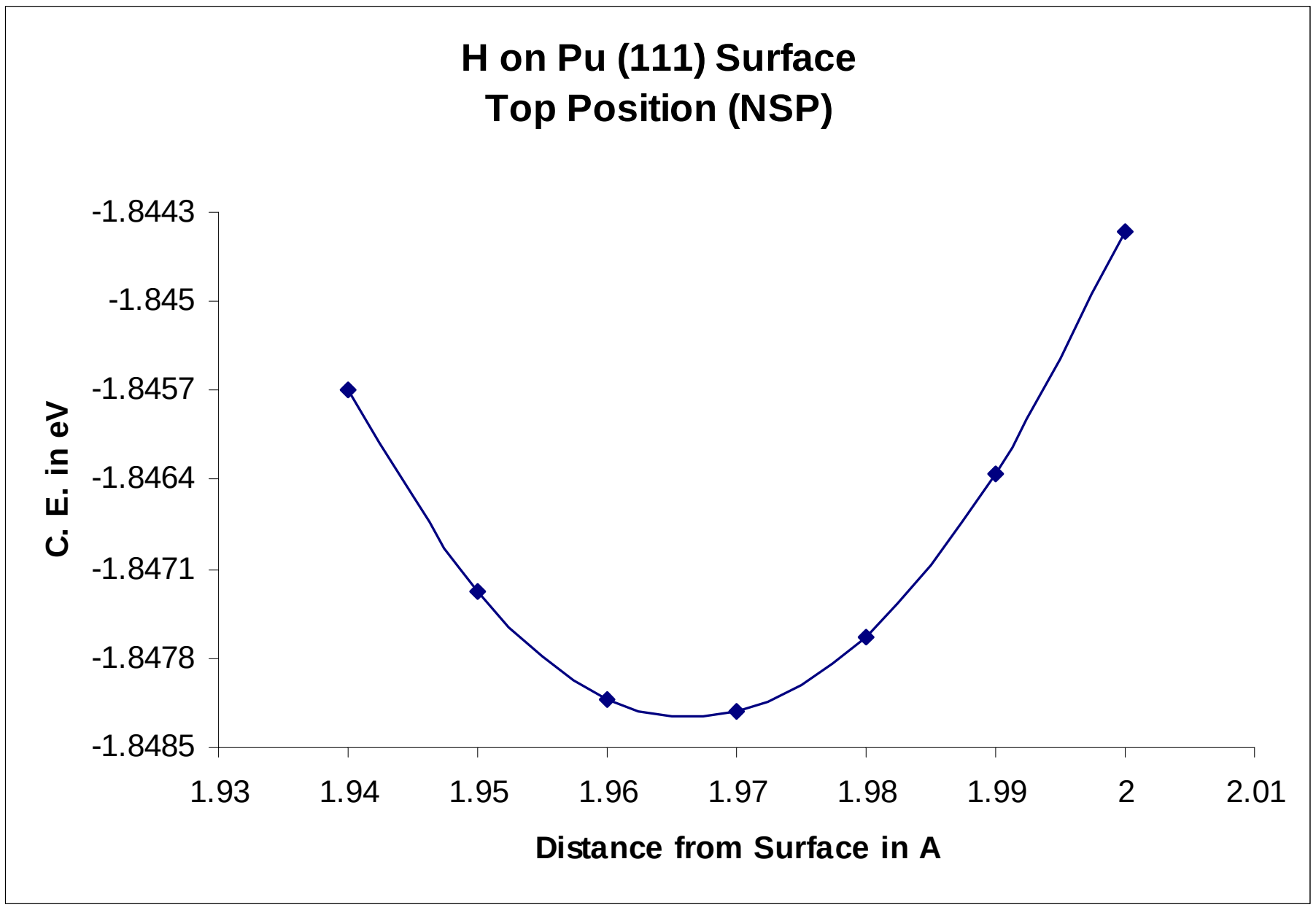

Figure 3(e). Chemisorption energy versus the hydrogen adatom distance from the Pu (111) surface in the top position without spin polarization.

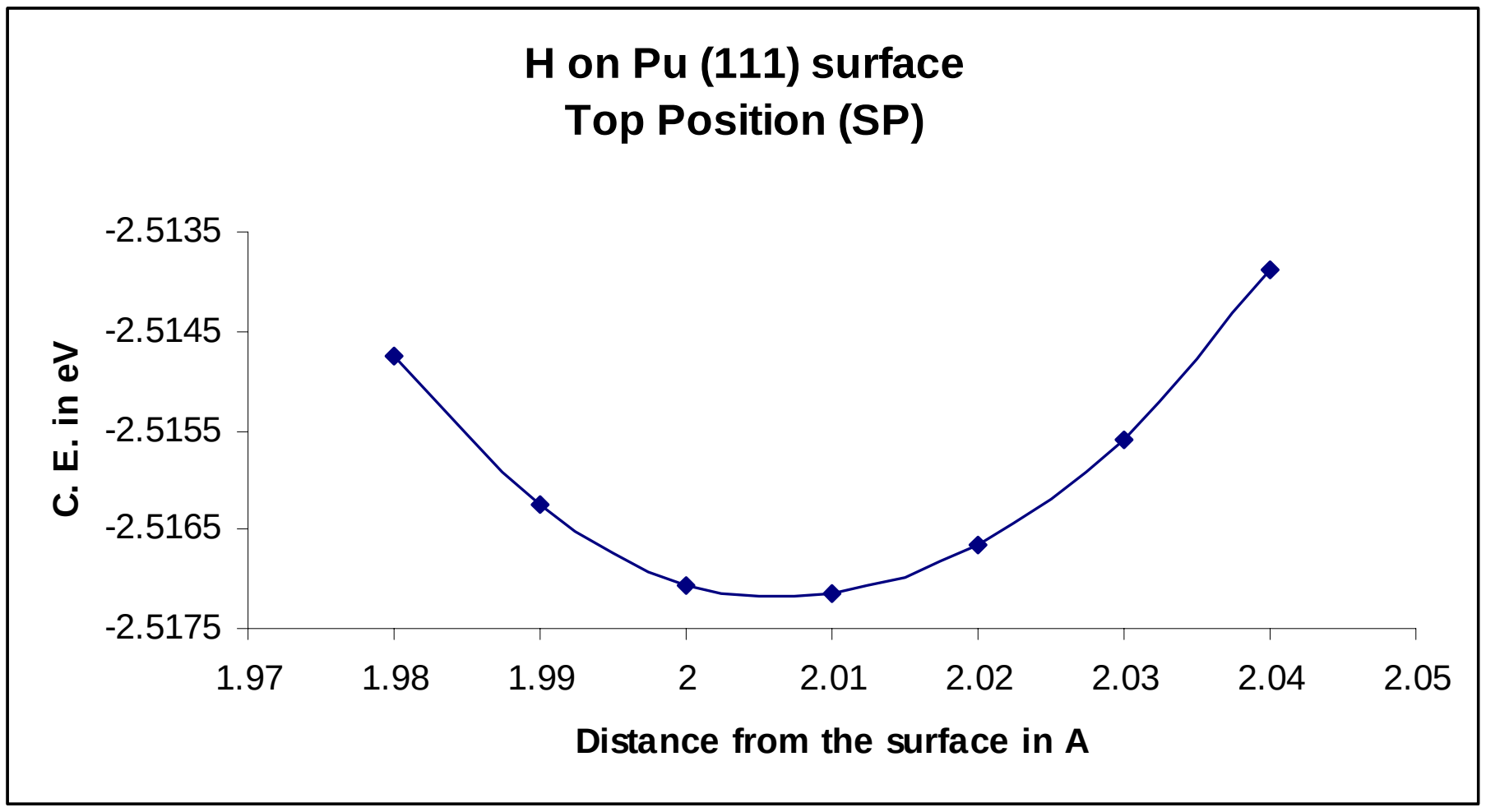

Figure 3(f). Chemisorption energy versus the hydrogen adatom distance from the Pu (111) surface in the top position with spin polarization.

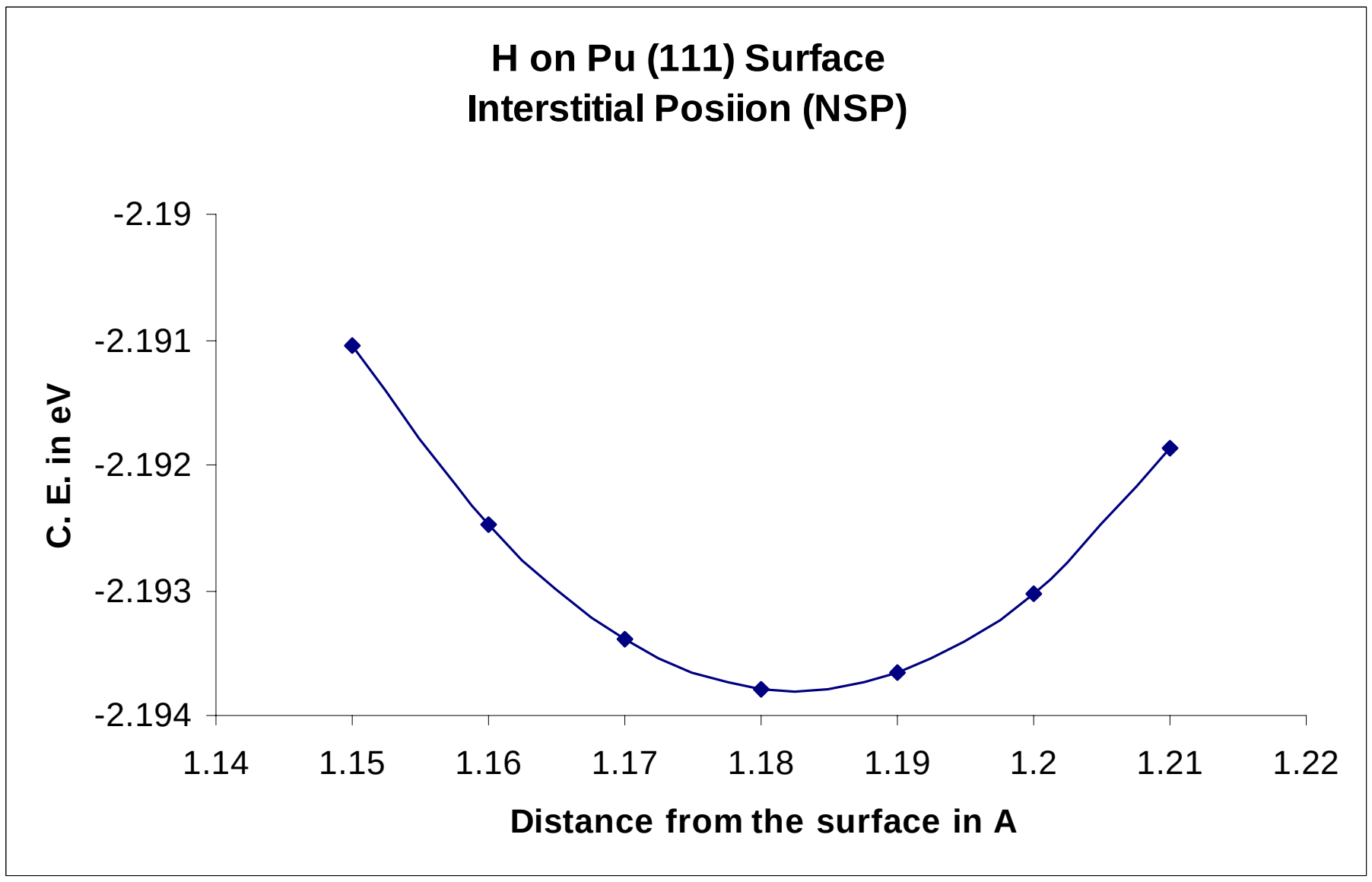

Figure 3(g). Chemisorption energy versus the hydrogen adatom distance from the Pu (111) surface in the interstitial position without spin polarization.

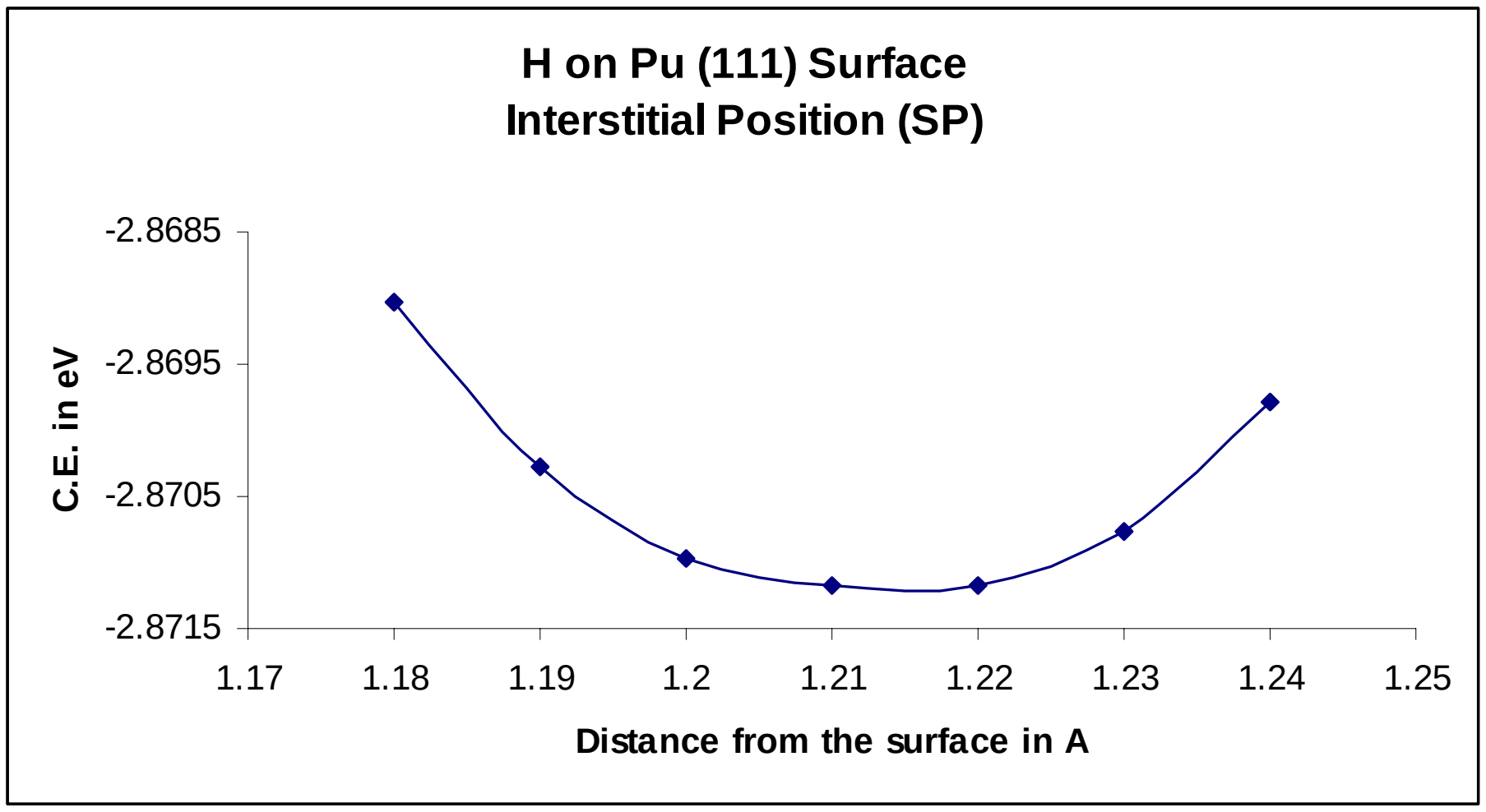

Figure 3(h). Chemisorption energy versus the hydrogen adatom distance from the Pu (111) surface in the interstitial position with spin polarization.